%% file: CorrelatedDoublyDirtyMAC_ver3.tex
\def\jolabels{0}
\def\numcolumns{1}

\if\numcolumns2
\documentclass[10pt,conference,twocolumn]{IEEEtran}
\else \if\numcolumns1
\documentclass[11pt,journal,onecolumn]{IEEEtran}
\renewcommand{\baselinestretch}{1.48}
\else 
 \typein{*** Press Enter ***} \stop \fi \fi

\renewcommand{\QED}{\QEDopen}

\usepackage{epsfig}
\usepackage{myamsthm}
\theoremstyle{plain}

\usepackage{color}
\usepackage{jolabels}

\input{j_shorts.tex}
\usepackage{amsmath}
\usepackage{fancybox}
\usepackage{latexsym}

\usepackage{graphicx}
\usepackage{amsfonts}
\usepackage{amssymb}
\usepackage{enumerate}

\newfont{\boldlarge}{msbm10 scaled 1100}

\newlength{\myVSpace}
\renewcommand{\theenumi}{{\bf\alph{enumi}}}
\makeatother

\begin{document}

\renewcommand{\textfraction}{0}
\title{\bf Technical Report:\\
Achievable Rates for the MAC with Correlated Channel-State
Information}

\author{
\large Tal Philosof, Ram Zamir and Uri Erez\\
\normalsize Dept. of Electrical Engineering - Systems, Tel-Aviv
University\\
\normalsize Tel-Aviv 69978, ISRAEL\\
{\em talp,zamir,uri@eng.tau.ac.il}\\
 }

\maketitle

\begin{abstract}
In this paper we provide an achievable rate region for the
discrete memoryless multiple access channel with correlated state
information known non-causally at the encoders using a random
binning technique. This result is a generalization of the random
binning technique used by Gel'fand and Pinsker for the problem
with non-causal channel state information at the encoder in point
to point communication.
\end{abstract}


\section{The Problem Setup}
We consider a discrete memoryless multiple access channel (MAC)
with two correlated states each known by one of the encoders.
Specifically, we assume the following model:
\begin{eqnarray}
P(y|x_1,x_2,s_1,s_2) \ \  \mbox{and} \ \ P(s_1,s_2),
\label{eq:GeneralModel}
\end{eqnarray}
where $s_1\in\mathcal{S}_1$ and $s_2\in\mathcal{S}_2$ are known
non-causally at encoder $1$ and encoder $2$, respectively. The
channel inputs are $x_1\in\mathcal{X}_1$ and
$x_2\in\mathcal{X}_2$, and the channel output is
$y\in\mathcal{Y}$. The memoryless channel implies that
\begin{align}
P(\mathbf{y}|\mathbf{x}_1,\mathbf{x}_2,\mathbf{s}_1,\mathbf{s}_2)=\prod_{i=1}^{n}P(y_i|{x_1}_i,{x_2}_i,{s_1}_i,{s_2}_i).
\end{align}

The first user transmits the message $m_1\in\{1,\ldots,M_1\}$, and
the second user transmits the message $m_2\in\{1,\ldots,M_2\}$,
where $m_1$ and $m_2$ are independent random variables with
uniform distributions, and $M_1=2^{nR_1}$, $M_2=2^{nR_2}$. The
first encoder observes the channel state information $S_1$
non-causally and generates the transmitted codeword
\begin{align}
\phi_1:\{1,\ldots,M_1\}\times
\mathcal{S}_1^n\rightarrow\mathcal{X}_1^n.
\end{align}
In the same way, the second encoder generates the transmitted
codeword
\begin{align}
\phi_2:\{1,\ldots,M_2\}\times
\mathcal{S}_2^n\rightarrow\mathcal{X}_2^n.
\end{align}
The decoder uses the following mapping to reconstruct the
transmitted messages
\begin{align}
\psi:\mathcal{Y}^n\rightarrow\{1,\ldots,M_1\}\times\{1,\ldots,M_2\},
\end{align}
i.e., $(\hat{m}_1,\hat{m}_2)=\psi(\mathbf{y})$. The error
probability is defined as
\begin{align}
P_e^{(n)}\triangleq\Pr\Big(\psi(\mathbf{y}) \neq (m_1,m_2) \Big).
\end{align}

\section{Main Result}
In the following theorem we provide an inner bound for the
capacity region of \eqref{eq:GeneralModel} which is derived using
a generalization of the random binning technique
\cite{GelfandPinsker80}.
\begin{theorem}\label{eq:InnerBound}
An inner bound for the capacity region of \eqref{eq:GeneralModel}
is given by
\begin{align}
\mathcal{R}\triangleq cl\;conv\; \Bigg\{(R_1,R_2):\;
&R_1\leq I(U;Y|V)-I(U;S_1|V)\nonumber\\
&R_2\leq I(V;Y|U)-I(V;S_2|U)\label{eq:RateRegionCorrelatedStates}\\
&R_1+R_2\leq I(U,V;Y)-I(U,V;S_1,S_2)\nonumber\\
&\text{for some admissible pair}\;(U,V)\qquad\qquad
\Bigg\}\nonumber
\end{align}
where the admissible pairs satisfy:
\begin{align}
P(U,V,X_1,X_2,S_1,S_2,Y)=P(S_1,S_2)P(U,X_1|S_1)P(V,X_2|S_2)P(Y|X_1,X_2,S_1,S_2).
\end{align}
\end{theorem}
The theorem implies that the following two Markov chains are
satisfied:
\begin{align}
&(U,X_1)\leftrightarrow S_1 \leftrightarrow S_2 \leftrightarrow (V,X_2)\\
&(U,V)\leftrightarrow (X_1,X_2,S_1,S_2) \leftrightarrow Y.
\end{align}
\begin{proof}
We denote the set of $\epsilon$-typical of two $n$-sequences
$\mathbf{a}$ and $\mathbf{b}$ where $a_i\in A, b_i\in B$ for
$i=1,\ldots,n$ by $A_{\epsilon}^{(n)}(A,B)$ (we use the same
notation as in \cite{CoverBook}).

Fix the distributions $P(U,X_1|S_1)$ and $P(V,X_2|S_2)$. Calculate
the marginal distributions $P(U)$ and $P(V)$.
\begin{itemize}
\item \textit{Codebooks generation}: Let
\begin{align}
J_1&=2^{n[I(U;S_1)+4\epsilon]}\\
J_2&=2^{n[I(V;S_2)+4\epsilon]}.
\end{align}
\newline
\textit{Codebook $1$}: Generate $2^{n(J_1+R_1)}$ of independent
$\mathbf{u}_k$ sequences of length $n$, generating each element
i.i.d according to distribution $\prod_{i=1}^n P(u_i)$, and
distribute these sequences randomly among $M_1$ bins where each
bin has $2^{nJ_1}$ sequences.
\newline
\textit{Codebook $2$}: Generate $2^{n(J_2+R_2)}$ of independent
$\mathbf{v}_k$ sequences of length $n$, generating each element
i.i.d according to distribution $\prod_{i=1}^n P(v_i)$, and
distribute these sequences randomly among $M_2$ bins where each
bin has $2^{nJ_2}$ sequences.

\item \textit{Encoder of user $1$}: Given the state sequence
$\mathbf{s}_1$ and the message $m_1$, search in bin $m_1$ of
codebook $1$ for a $\mathbf{u}$ sequence such that
$(\mathbf{u},\mathbf{s}_1)\in A_{\epsilon}^n(U,S_1)$. Send
$\mathbf{x}_1$ which is jointly typical with $\mathbf{u}$ and
$\mathbf{s}_1$, i.e., $(\mathbf{u},\mathbf{s}_1,\mathbf{x}_1)\in
A_{\epsilon}^n(U,S_1,X_1)$

\item \textit{Encoder of user $2$}: Given the state sequence
$\mathbf{s}_2$ and the message $m_2$, search in bin $m_2$ of
codebook $2$ for a $\mathbf{v}$ sequence such that
$(\mathbf{v},\mathbf{s}_2)\in A_{\epsilon}^n(V,S_2)$. Send
$\mathbf{x}_2$ which is jointly typical with $\mathbf{v}$ and
$\mathbf{s}_2$, i.e., $(\mathbf{v},\mathbf{s}_2,\mathbf{x}_2)\in
A_{\epsilon}^n(U,S_2,X_2)$

\item \textit{Decoder}: Given the received vector $\mathbf{y}$,
search for unique sequences $\mathbf{u}$ and $\mathbf{v}$ such
that $(\mathbf{u},\mathbf{v},\mathbf{y})\in
A_{\epsilon}^n(U,V,Y)$.

\end{itemize}

\textit{Analysis of the error probability}: The error probability
is given by
\begin{align*}
P_e^{(n)}&=\sum_{\mathbf{s}_1,\mathbf{s}_2\notin
A_{\epsilon}^{(n)}(S_1,S_2)}P_{S_1^n,S_2^n}(\mathbf{s}_1,\mathbf{s}_2)
+\sum_{\mathbf{s}_1,\mathbf{s}_2\in
A_{\epsilon}^{(n)}(S_1,S_2)}P_{S_1^n,S_2^n}(\mathbf{s}_1,\mathbf{s}_2)P(e|\mathbf{s}_1,\mathbf{s}_2)\\
&\leq \epsilon+\sum_{\mathbf{s}_1,\mathbf{s}_2\in
A_{\epsilon}^{(n)}(S_1,S_2)}P_{S_1^n,S_2^n}(\mathbf{s}_1,\mathbf{s}_2)P(e|\mathbf{s}_1,\mathbf{s}_2),
\end{align*}
where the inequality follows the asymptotic equipartition property
(AEP) \cite{CoverBook}.  Hence, we need to evaluate only the
second term. We define the following error events for specific
sate sequences $(\mathbf{s}_1,\mathbf{s}_2)$ :
\begin{itemize}
\item $E_1(\mathbf{s}_1,m_1)=\{\nexists\;j_1,\;1\leq j_1\leq
J_1\;:\;(\mathbf{u}_{m_1,j_1},\mathbf{s}_1)\in
A_{\epsilon}^{(n)}(U,S_1)\}$.

\item $E_2(\mathbf{s}_2,m_2)=\{\nexists\;j_2,\;1\leq j_2\leq
J_2\;:\;(\mathbf{v}_{m_2,j_2},\mathbf{s}_2)\in
A_{\epsilon}^{(n)}(V,S_2)\}$.

\item
$E_3(\mathbf{s}_1,\mathbf{s}_2,m_1,m_2)=\{(\mathbf{u}_{m_1,j_1(\mathbf{s_1},m_1)},\mathbf{v}_{m_2,j_2(\mathbf{s_2},m_2)},\mathbf{s}_1,\mathbf{s}_2)\notin
A_{\epsilon}^{(n)}(U,V,S_1,S_2)\}$.

\item
$E_4(\mathbf{s}_1,\mathbf{s}_2,m_1,m_2)=\{(\mathbf{u}_{m_1,j_1(\mathbf{s_1},m_1)},\mathbf{v}_{m_2,j_2(\mathbf{s_2},m_2)},\mathbf{y})\notin
A_{\epsilon}^{(n)}(U,V,Y)\}$.

\item $E_5(\mathbf{s}_1,\mathbf{s}_2,m_1,m_2)=\{\exists
\mathbf{u}_{m'_1,j_1}: m_1\neq m'_1,
(\mathbf{u}_{m'_1,j_1},\mathbf{v}_{m_2,j_2},\mathbf{y})\in
A_{\epsilon}^{(n)}(U,V,Y)\}$.

\item $E_6(\mathbf{s}_1,\mathbf{s}_2,m_1,m_2)=\{\exists
\mathbf{v}_{m'_2,j_2}: m_2\neq m'_2,
(\mathbf{u}_{m_1,j_1},\mathbf{v}_{m'_2,j_2},\mathbf{y})\in
A_{\epsilon}^{(n)}(U,V,Y)\}$.

\item $E_7(\mathbf{s}_1,\mathbf{s}_2,m_1,m_2)=\{\exists
\mathbf{u}_{m'_1,j_1},\mathbf{v}_{m'_2,j_2}: m_1\neq m'_1, m_2\neq
m'_2, (\mathbf{u}_{m'_1,j_1},\mathbf{v}_{m'_2,j_2},\mathbf{y})\in
A_{\epsilon}^{(n)}(U,V,Y)\}$.

\end{itemize}
Then by union bound, the error probability is upper bounded by
\begin{align}
P_e^{(n)}\leq&\; \epsilon+\frac{1}{M_1M_2}\sum_{m_1,m_2}P_{S_1^n,S_2^n}(\mathbf{s}_1,\mathbf{s}_2)\Big[\Pr(E_1(\mathbf{s}_1,m_1))+\Pr(E_2(\mathbf{s}_2,m_2))\nonumber\\
&+\Pr(E_3(\mathbf{s}_1,\mathbf{s}_2,m_1,m_2)|\overline{E}_1(\mathbf{s}_1,m_1),\overline{E}_2(\mathbf{s}_2,m_2))\nonumber\\
&+\Pr(E_4(\mathbf{s}_1,\mathbf{s}_2,m_1,m_2)|\overline{E}_3(\mathbf{s}_1\mathbf{s}_2,m_1,m_2))\nonumber\\
&+\Pr(E_5(\mathbf{s}_1,\mathbf{s}_2,m_1,m_2)|\overline{E}_3(\mathbf{s}_1\mathbf{s}_2,m_1,m_2))\\
&+\Pr(E_6(\mathbf{s}_1,\mathbf{s}_2,m_1,m_2)|\overline{E}_3(\mathbf{s}_1\mathbf{s}_2,m_1,m_2))\nonumber\\
&+\Pr(E_7(\mathbf{s}_1,\mathbf{s}_2,m_1,m_2)|\overline{E}_3(\mathbf{s}_1,\mathbf{s}_2,m_1,m_2))\Big]\nonumber
\end{align}

We now evaluate the probability of each error events. For
independent $\mathbf{u}$ and $\mathbf{s}_1$ the probability that
$(\mathbf{u},\mathbf{s}_1)\in A_{\epsilon}^{n}(U,S_1)$ is bounded
below by
\begin{align*}
\Pr\left((\mathbf{u},\mathbf{s}_1)\in A_{\epsilon}^{n}(U,S_1)\right) &= \sum_{(\mathbf{u},\mathbf{s}_1)\in A_{\epsilon}^{n}(U,S_1)}P(\mathbf{u})P(\mathbf{s}_1)\\
&\geq
|A_{\epsilon}^{n}(U,S_1))|2^{-n[H(U)+\epsilon]}2^{-n[H(S_1)+\epsilon]}\\
&\geq 2^{n[H(U,S_1)-\epsilon]}
2^{-n[H(U)+\epsilon]}2^{-n[H(S_1)+\epsilon]}\\
&=2^{-n[H(U)+H(S_1)-H(U,S_1)+3\epsilon]}\\
&=2^{-n[I(U;S_1)+3\epsilon]}
\end{align*}
Hence, we have that
\begin{align}
\Pr(E_1(\mathbf{s}_1,m_1))&\leq
\left[1-2^{-n[I(U;S_1)+3\epsilon]}\right]^{J_1}\\
&\leq \exp\left(-J_12^{-n[I(U;S_1)+3\epsilon]}\right)\label{eq:10}\\
&= \exp(-2^{n\epsilon}),
\end{align}
where \eqref{eq:10} follows since $1-x\leq \exp(-x)$. Hence, this
term decays to zero as $n\rightarrow \infty$. In the same way $
\Pr(E_2(\mathbf{s}_2,m_2))$ goes to zero as $n\rightarrow \infty$.

Provided that $E_1(\mathbf{s}_1,m_1)$ and $E_2(\mathbf{s}_2,m_2)$
have not occurred, i.e., $(\mathbf{u}_{m_1,j_1},\mathbf{s}_1)\in
A_{\epsilon}^{(n)}(U,S_1)$ and
$(\mathbf{v}_{m_2,j_2},\mathbf{s}_2)\in
A_{\epsilon}^{(n)}(V,S_2)$, from Markov Lemma \cite{CoverBook} we
have that
\begin{align}
\Pr((\mathbf{u},\mathbf{v},\mathbf{s}_1,\mathbf{s}_2)\in
A_{\epsilon}^{(n)}(U,V,S_1,S_2)|(\mathbf{u},\mathbf{s}_1)\in
A_{\epsilon}^{(n)}(U,S_1),(\mathbf{v},\mathbf{s}_2)\in
A_{\epsilon}^{(n)}(V,S_2))\geq 1-\epsilon
\end{align}
where the typical set $A_{\epsilon}^{(n)}(U,V,S_1,S_2)$ is
associated with the joint distribution
\begin{align}
P(U,V,S_1,S_2) = P(S_1,S_2)P(U|S_1)P(V|S_2)\label{eq:JointUSV}.
\end{align}
Hence, we have that
\begin{align}
\Pr(E_3(\mathbf{s}_1,\mathbf{s}_2,m_1,m_2)|\overline{E}_1(\mathbf{s}_1,m_1),\overline{E}_2(\mathbf{s}_2,m_2)\leq
\epsilon
\end{align}
In fact, we have (with high probability) that the sequences
$(\mathbf{u}_{m_1,j_1},\mathbf{v}_{m_2,j_2},\mathbf{s}_1,\mathbf{s}_2)$
generated using the joint distribution \eqref{eq:JointUSV}, which
is equivalent to the Markov chain $U \leftrightarrow S_1
\leftrightarrow S_2 \leftrightarrow V$.
%
%

Provided that $E_3(\mathbf{s}_1,\mathbf{s}_2,m_1,m_2)$ has not
occurred, from the AEP we have that
\begin{align*}
&\Pr(E_4(\mathbf{s}_1,\mathbf{s}_2,m_1,m_2)|\overline{E}_3(\mathbf{s}_1,\mathbf{s}_2,m_1,m_2))\\
&=\Pr((\mathbf{u},\mathbf{v},\mathbf{y})\notin
A_{\epsilon}^{(n)}(U,V,Y)|(\mathbf{u},\mathbf{s}_1)\in
A_{\epsilon}^{(n)}(U,S_1),(\mathbf{v},\mathbf{s}_2)\in
A_{\epsilon}^{(n)}(V,S_2))\\
&\leq \epsilon.
\end{align*}
Likewise, we have that
\begin{align}
&\Pr(E_5(\mathbf{s}_1,\mathbf{s}_2,m_1,m_2)|\overline{E}_3(\mathbf{s}_1,\mathbf{s}_2,m_1,m_2))\\
& \leq M_1 J_1
\Pr((\mathbf{u}_{m'_1,j_1},\mathbf{v}_{m_2,j_2},\mathbf{y})\in
A_{\epsilon}^{(n)}(U,V,Y))\\
&= M_1 J_1 \sum_{(\mathbf{u},\mathbf{v},\mathbf{y})\in
A_{\epsilon}^{(n)}(U,V,Y)}p(\mathbf{u})p(\mathbf{v},\mathbf{y})\\
&\leq M_1 J_1 |A_{\epsilon}^{(n)}(U,V,Y)|
2^{-n[H(U)-\epsilon]}2^{-n[H(V,Y)-\epsilon]}\label{eq:100}\\
&\leq 2^{nR_1}2^{n[I(U;S_1)+4\epsilon]}2^{-n[H(U)+H(V,Y)-H(U,V,Y)-3\epsilon]}\label{eq:110}\\
&= 2^{nR_1}2^{n[I(U;S_1)+4\epsilon]}2^{-n[I(U;V,Y)-3\epsilon]}\\
&= 2^{nR_1}2^{n[I(U;S_1)+4\epsilon]}2^{-n[I(U;V)+I(U;Y|V)-3\epsilon]}\label{eq:120}\\
&= 2^{nR_1}2^{-n[I(U;Y|V)+I(U;V)-I(U;S_1)-7\epsilon]}\\
&= 2^{nR_1}2^{-n[I(U;Y|V)-H(U|V)+H(U|S_1)-7\epsilon]}\\
&= 2^{nR_1}2^{-n[I(U;Y|V)-H(U|V)+H(U|S_1,V)-7\epsilon]}\label{eq:130}\\
&= 2^{nR_1}2^{-n[I(U;Y|V)-I(U;S_1|V)-7\epsilon]}\label{eq:200},
\end{align}
where \eqref{eq:100} and \eqref{eq:110} follow from AEP;
\eqref{eq:120} follows from the chain rule for mutual information;
\eqref{eq:130} follows from the Markov chain $U\leftrightarrow S_1
\leftrightarrow V$. In the same way, it can be shown that
\begin{align}
&\Pr(E_6(\mathbf{s}_1,\mathbf{s}_2,m_1,m_2)|\overline{E}_3(\mathbf{s}_1,\mathbf{s}_2,m_1,m_2))
\leq 2^{nR_2}2^{-n[I(V;Y|U)-I(V;S_2|U)-7\epsilon]}\label{eq:300}
\end{align}
Furthermore,
\begin{align}
&\Pr(E_7(\mathbf{s}_1,\mathbf{s}_2,m_1,m_2)|\overline{E}_3(\mathbf{s}_1,\mathbf{s}_2,m_1,m_2))\\
& \leq M_1M_2 J_1J_2
\Pr((\mathbf{u}_{m'_1,j_1},\mathbf{v}_{m'_2,j_2},\mathbf{y})\in
A_{\epsilon}^{(n)}(U,V,Y))\\
&= M_1M_2 J_1J_2 \sum_{(\mathbf{u},\mathbf{v},\mathbf{y})\in
A_{\epsilon}^{(n)}(U,V,Y)}P(\mathbf{u})P(\mathbf{v})P(\mathbf{y})\\
&\leq M_1M_2 J_1J_2 |A_{\epsilon}^{(n)}(U,V,Y)|
2^{-n[H(U)-\epsilon]}2^{-n[H(V)-\epsilon]}2^{-n[H(Y)-\epsilon]}\label{eq:310}\\
&= M_1M_2 J_1J_2 2^{n[H(U,V,Y)-\epsilon]}
2^{-n[H(U)+H(V)+H(Y)-3\epsilon]}\label{eq:320}\\
&\leq 2^{n[R_1+R_2]}2^{n[I(U;S_1)+4\epsilon]}2^{n[I(V;S_2)+4\epsilon]}2^{-n[H(U)+H(V)+H(Y)-H(U,V,Y)-3\epsilon]}\\
&= 2^{n[R_1+R_2]}2^{n[I(U;S_1)+I(V;S_2)+8\epsilon]}2^{-n[I(U,V;Y)+I(U;V)-3\epsilon]}\\
&= 2^{n[R_1+R_2]}2^{-n[I(U,V;Y)-I(U;S_1)-I(V;S_2)+I(U;V)-11\epsilon]}\\
&= 2^{n[R_1+R_2]}2^{-n[I(U,V;Y)+H(U|S_1)-H(U|V_1)-I(V;S_2)-11\epsilon]}\\
&= 2^{n[R_1+R_2]}2^{-n[I(U,V;Y)+H(U|S_1,S_2,V)-H(U|V)-I(V;S_2)-11\epsilon]}\label{eq:330}\\
&= 2^{n[R_1+R_2]}2^{-n[I(U,V;Y)-I(U;S_1,S_2|V)-I(V;S_1,S_2)-11\epsilon]}\\
&=
2^{n[R_1+R_2]}2^{-n[I(U,V;Y)-I(U,V;S_1,S_2)-11\epsilon]}\label{eq:400}
\end{align}
where \eqref{eq:310} and \eqref{eq:320} follow from AEP;
\eqref{eq:330} follows from the Markov chain $U\leftrightarrow S_1
\leftrightarrow S_2 \leftrightarrow V$; \eqref{eq:400} follows
from the chain rule for mutual information.

The theorem follows from \eqref{eq:200}, \eqref{eq:300},
\eqref{eq:400}, since for any arbitrary $\epsilon>0$ the
conditions in \eqref{eq:RateRegionCorrelatedStates} imply that
$P_e^{(n)}\rightarrow 0$ as $n\rightarrow \infty$.
\end{proof}

\section{Special Cases}
We consider now two special cases of the memoryless MAC with
correlated state information known non-causally at the encoders.
The first case is for $S_1=S_2$, i.e., the relation between the
states is deterministic. The second case is for independent
states.
\begin{enumerate}[I.]
\item \emph{Single state}: in this case we have single state which
is known to both encoders, i.e., $S=S_1=S_2$, the achievable rate
region is given by
\begin{align}
\mathcal{R}\triangleq cl\;conv\; \Bigg\{(R_1,R_2):\;
&R_1\leq I(U;Y|V)-I(U;S|V)\nonumber\\
&R_2\leq I(V;Y|U)-I(V;S|U)\label{eq:RegionEqualS}\\
&R_1+R_2\leq I(U,V;Y)-I(U,V;S)\nonumber\\
&\text{for some admissible pair}\;(U,V)\qquad\qquad
\Bigg\}\nonumber
\end{align}
where the admissible pairs satisfy: $(U,X_1)\leftrightarrow S
\leftrightarrow (V,X_2)$, and $(U,V)\leftrightarrow (X_1,X_2,S)
\leftrightarrow Y$.

The Gaussian case of single interference is given by
\begin{align}
Y=X_1+X_2+S+Z,
\end{align}
where $Z\sim\mathcal{N}(0,N)$, the interference $S$ is known
non-causally to user $1$ and user $2$, and the power constraints
are $P_1$ and $P_2$ for user $1$ and user $2$, respectively. This
model was considered by Gel'fand and Pinsker
\cite{GelfandPinskerISIT83}. It was shown that the capacity region
is equal to \emph{clean} MAC, i.e., for the case that $S=0$. In
this case, the region in \eqref{eq:RegionEqualS} concises with the
clean MAC region \cite{CoverBook}.

\item \emph{Independent states}: for the case that $S_1$ and $S_2$
are independent, the achievable region becomes
\begin{align}
\mathcal{R}\triangleq cl\;conv\; \Bigg\{(R_1,R_2):\;
&R_1\leq I(U;Y|V)-I(U;S_1)\nonumber\\
&R_2\leq I(V;Y|U)-I(V;S_2)\label{eq:RateRegionIndepStates}\\
&R_1+R_2\leq I(U,V;Y)-I(U;S_1)-I(V,S_2)\nonumber\\
&\text{for some admissible pair}\;(U,V)\qquad\qquad
\Bigg\}\nonumber
\end{align}
where the admissible pairs satisfy: $(U,S_1,X_1)$ is independent
of $(V,S_2,X_2)$, and $(U,V)\leftrightarrow (X_1,X_2,S_1,S_2)
\leftrightarrow Y$. The case with independent channel states was
originally considered in \cite{Jafar06}, which also introduces the
rate region in  \eqref{eq:RateRegionIndepStates}.

\end{enumerate}

\bibliographystyle{IEEEtran}
\bibliography{../../../Bibliography/mybib}

\end{document}

%% file: j_shorts.tex
\newtheorem{theorem}{Theorem}

\newcommand{\rem}[1]{}

\newcommand{\beq}{\begin{equation}}
\newcommand{\eeq}{\end{equation}}